# ZnO UV photodetector with controllable quality factor and photosensitivity

L C Campos[a,b], M H D Guimarães[c], A M B Goncalves, S de Oliveira and R G Lacerda

*Departamento de Física, Universidade Federal de Minas Gerais, Belo Horizonte, MG, CEP: 30123-970, Brazil*

ZnO nanowires have an enormous potential for applications as ultra-violet (UV) photodetectors. Their mechanism of photocurrent generation is intimately related with the presence of surface states where considerable efforts, such as surface chemical modifications, have been pursued to improve their photodetection capabilities. In this work, we report a step further in this direction demonstrating that the photosensitivity and quality factor ($Q$ factor) of the photodetector are entirely tunable by an applied gate voltage. This mechanism enables UV photodetection selectivity ranging from wavelengths from tens of nanometers (full width at half maximum - FWHM) down to a narrow detection of *3 nm*. Such control paves the way for novel applications, especially related to the detection of elements that have very sharp luminescence.

---

[a] L. C. Campos is currently at Department of Physics, Massachusetts Institute of Technology, Cambridge, Massachusetts 02139, USA
[b] Electronic mail: lccampos@fisica.ufmg.br
[c] M. H. D. Guimarães is currently at Physics of Nanodevices, Zernike Institute for Advanced Materials, University of Groningen, The Netherlands

## I. INTRODUCTION

Semiconductor Nanowires (NW) with novel functionalities have attracted great interest for electronic applications such as piezoelectronics energy conversion devices[1–4], as well as sensors[5–9], memory[10] and optical devices[11–16]. Among such systems, ZnO has always stood out due to its attractive wide band gap (*3.37 eV* at room temperature) with a high exciton binding energy (*60 meV*), which makes it appropriate for fabrication of short wavelength optoelectronic applications, such as ultra-violet (UV) detectors, and UV and blue light emitting devices[11,17,18]. Despite the extensive research and the achieved progress the integration of ZnO nanowires as a complete optoelectronic device is still a challenge. In the past decade some important advances have been made in this direction. For instance surface chemical funcionalization has been used to improve the ZnO photo-detecting performance such as gain and response time[19,20]. This occurs because the surface trap states affect the process of recombination and diffusion in ZnO[11,12,16]. In addition, the development of ZnO phototransistors has provided an extra parameter for the photocurrent switching behavior (ON/OFF), where gate voltage is used to add or suppress charges in the nanowire[10,21,22]. In this context, Kim *et al.*[23] have shown that a ZnO UV photodetector has its photosensitivity and time response dependent on the gate voltage, which points to a tunable way of controlling the photodetector properties. Here we show for the first time that the gate voltage can be used to externally manipulate the *Q* factor of a ZnO NW photodetector. This control is due to the depletion or accumulation of charges within the nanowire by the gate voltage either when it is absorbing light or not. This means that when the intensity of the light is constant, the photocurrent generated has its maximum value at the energy gap of the ZnO, following its absorption/photo-luminescence spectrum[17,24]. Consequently, the magnitude of threshold gate voltage ($V_{th}$) necessary to deplete the charges is also the largest around the energy of the gap allowing us to use this parameter to squeeze the photodetection nearby this point. This approach leads to a controllable detection of UV wavelengths ranging from tens of nanometers full width at half-maximum (FWHM) down to a narrow detection of ~3 nm. Indeed, we also observed that the maximum photosensitivity of the ZnO photodetector is achieved when the gate voltage is settled as the $V_{th}$ of the phototransistor where light is no longer absorbed. We believe that the realization of photodetectors with improved quality factors has strategic importance and may ignite new applications such as the development of suitable detectors of hazardous airborne elements that have very sharp luminescence[25,26].

## II. SAMPLE FABRICATION

Zinc Oxide NWs are grown by a thermal oxidation process of Zn *99.9%* pure foil. Details of the growth conditions and analysis of the NW structure have been reported elsewhere[27]. Briefly, a Zn foil is inserted into an opened furnace at ambient air and heated at *500 °C* for *2* hours. As result of the Zn thermal oxidation, the Zn foil is covered by a "forest" of ZnO nanowires, with diameters ranging from a few nanometers to hundreds of nanometers (ZnO crystalline composition was identified by synchrotron X-ray diffraction and TEM)[27]. The nanowire dispersion on the $SiO_2$/Si substrate for device fabrication is performed in three steps. First, the foil is immersed in isopropyl alcohol and ultrasonicated for *2* seconds. Afterwards, the resulting solution is deposited on heavily doped Si substrate with a thermally grown *300 nm* $SiO_2$ film on the surface. The samples are then dried at *180 °C* on a hotplate for *2* minutes. Finally, Cr/Au (*5/200 nm*) contacts are deposited on the isolated NW using conventional optical lithography and metal deposition techniques. A (back) gate voltage is applied to the heavily doped Si substrate (Fig. 1(a)).

## III. RESULTS

The electrical behavior of the isolated ZnO nanowire device in dark conditions is summarized in Fig. 1(b). The constant resistance in the source-drain current ($I_{sd}$) versus source-drain voltage ($V_{sd}$) curves confirms ohmic metal - ZnO contact. Appling $V_{sd} = 1$ V and changing the gate voltage ($V_g$), we observe the expected n-type behavior of the ZnO nanowire and a corresponding ON/OFF ratio of the order of $10^3$ ($I_{sd}$ changes from *nA* to *μA*). In Fig. 2(a), the source-drain current versus gate voltage for different incident wavelengths is shown. Analyzing the curve for dark conditions (Fig. 2(a) black line), it can be seen that for voltages below the threshold voltage ($V_{th} = -13$ V) the current is suppressed indicating the depletion of charge carriers within the NW. The field effect mobility is $\mu = 35$ *cm²/Vs* and the carrier concentration at $V_g = 0$ is $n = 6\times10^{16}$ *cm⁻³*, which was calculated using $\mu = egL/cV_{sd}$ and $n = V_{th}C/eAL$ where $e$ is the elementary charge, $g$ is the transconductance, $L$ is the length of the wire, $V_{sd}$ is the source-drain voltage, $C$ the capacitance and $A$ the cross sectional area of the nanowire extracted by scanning electron microscopy images[28,29]. The mobility and charge densities obtained are in good agreement with typical values reported in the literature for ZnO nanowires[30–33]. To further characterize the device, we measure the $V_g$ dependence of the photocurrent for a range of wavelengths. The light source was a "Halogen Lamp", with a quasi-constant emission from *325 nm* to *475 nm*, which was passed through a monochromator and then focused on the nanowire. No photocurrent is observed for wavelengths larger than *380 nm* and a maximum occurs at an incidence wavelength of *371 nm*, which corresponds to the band gap of the ZnO nanowire, $E_g = 3.34$ *eV* (Fig. 2(b)). For this wavelength, the current flowing through the nanowire increases about *3* orders of magnitude with respect to the current in dark conditions. The carriers generated via photo absorption can be considered as a variation of the carrier concentration ($\Delta n$), in this quasi-equilibrium condition, subjected to a uniform electric field[34]. Therefore, in a phototransistor the $V_{th}$ proportionally varies with the density of carriers which strongly depends on the surface states, the environment, temperature and the incidence of light[35–37].

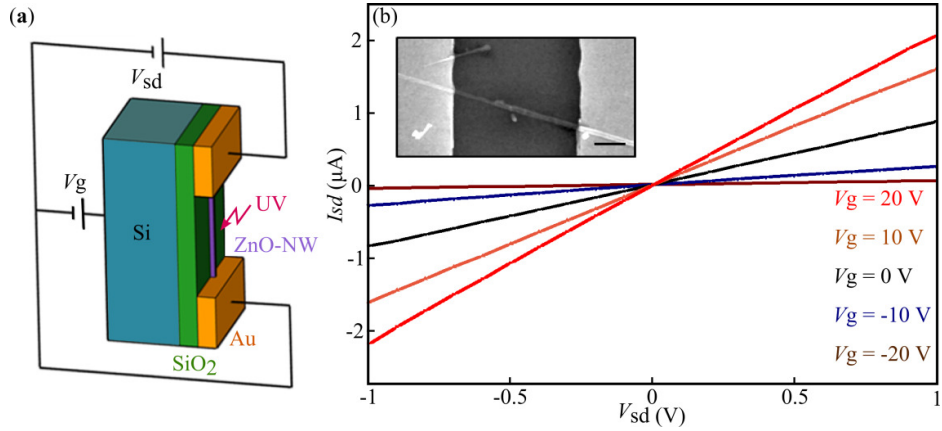

FIG. 1. (a) Schematic cartoon of the ZnO nanowire transistor. (b) $I_{sd}$ versus $V_{sd}$ as a for different values of $V_g$. The conductance of the nanowire increases for positive $V_g$ showing the n-type characteristic of the ZnO nanowire field effect transistor. Inset on figure 1(b) shows a scanning electron microscopy image of the ZnO nanowire device (scale bar *1 μm*).

When one measures the dependence of the $I_{sd}$ current as function of the gate voltage for different incident wavelengths (Fig. 2(a)), one can see that under photo-absorption, the photocurrent increases and $V_{th}$ shifts, indicating an increase of the total density of charges in the nanowire. Additionally, it is also important to point out that in the particular case of ZnO nanowires the photo absorption spectrum typically has a narrow peak near the band edge in the exciton absorption region[17,24]. As a consequence of that, $V_{th}$ will also have its maximum value at the photoabsorption peak, due to the highest increase in carrier density. Moreover, this direct dependence of the $V_{th}$ with the energy of the incident light opens a new application of ZnO phototransistors because the gate voltage can be used to make the device more selective to certain frequencies. The gate voltage works as an externally adjustable parameter that selects which range of wavelengths (around the band gap) will create a photocurrent, therefore making the photodetector more selective to frequencies.

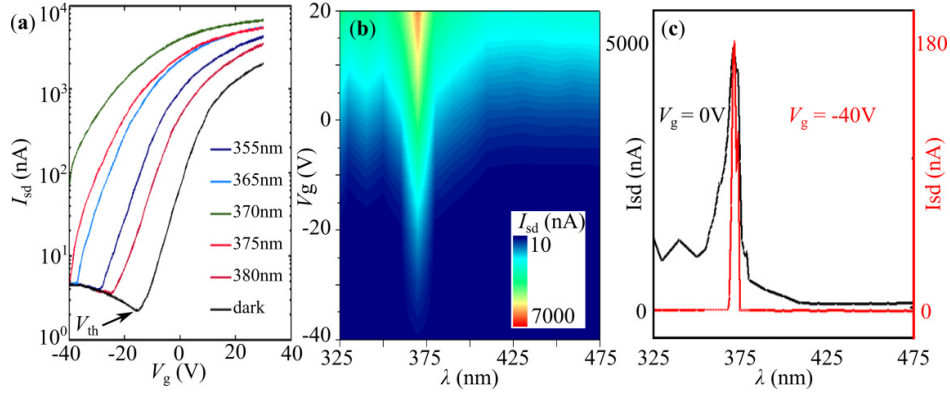

FIG. 2. (a) Source-drain current ($I_{sd}$) versus gate-voltage ($V_g$) for different values of incident wavelength ($\lambda$) for a fixed source-drain bias of $V_{sd} = 1\ V$. The $V_{th}$ increases in modulus with the incident wavelength up to a peak around *371 nm*. (b) Photocurrent as function of the wavelengths for different gate voltages ($V_{sd} = 1\ V$). (c) The photocurrent as a function of wavelengths for two fixed gate voltages ($V_g = 0\ V$ in black, and $V_g = -40\ V$ in red).

A careful analysis of Fig. 2(a) shows that for a fixed $V_g < V_{th}$ the nanowire will be depleted of charges unless the photoabsorption is strong enough to generate free carriers. This effect can be clearly observed in Fig. 2(b), where a photocurrent spectroscopy is shown as a function of the gate voltage. At negative gate voltages, the intensity of the current decreases as the photo-detection becomes restricted to wavelengths near *371 nm*. Fig. 2(c) shows a comparative spectrum of the photocurrents for $V_g = 0\ V$, which resembles the photo-absorption spectrum, and for $V_g = -40\ V$, where the photocurrent is generated only at wavelengths nearby *371 nm*. In fact, our results show that there is a linear relationship $(V_{th} - V_{dark}) \propto I_{ph}$ between the changes of the $V_{th}$ and the photocurrent, which is a direct consequence of the diffusive transport of the charges (not shown).

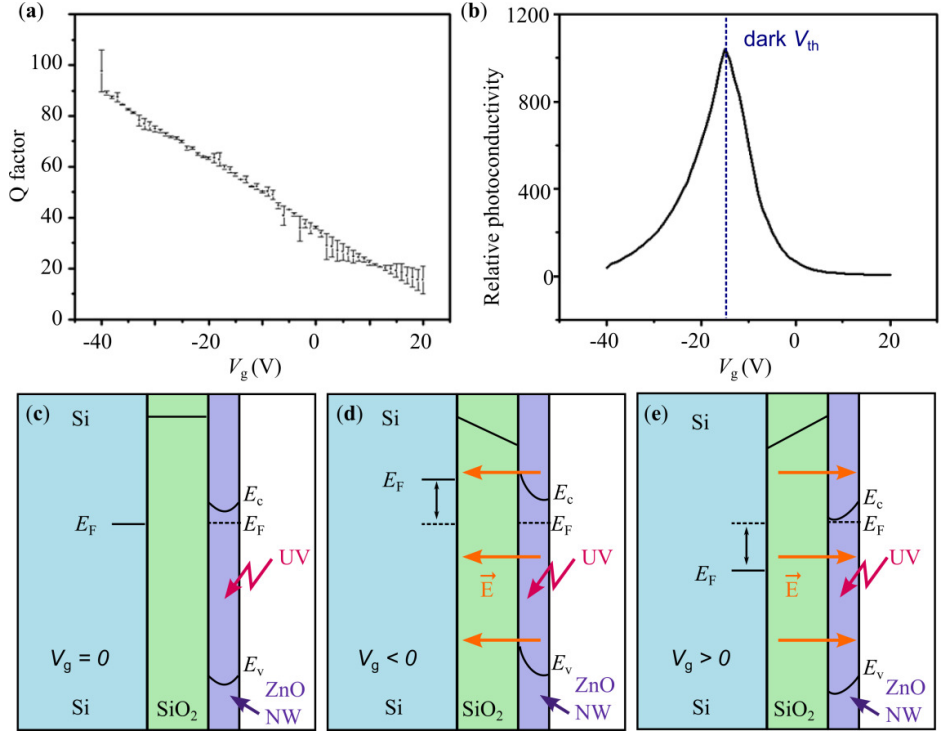

FIG. 3. (a) Quality factor and (b) the relative photoconductivity photosensitivity as a function $V_g$. UV exposure at (c) $V_g = 0\ V$, (d) $V_g < 0\ V$ and (e) $V_g > 0\ V$.

The effect of the gate voltage on the general performance of the photodetector is clearly noticeable when we evaluate the quality factor (Q factor) and relative photoconductivity (Figure 3(a) and 3(b)). The Q factor (Fig. 3(a)), defined by $Q=\Delta\lambda/\lambda$, linearly decreases with $V_g$, which confirms that for more negative gate voltages the photodetector becomes more selective. Here we consider that $\Delta\lambda$ corresponds to the wavelength absorption width (FWHM) around $\lambda = 371\ nm$. Also, the relative photoconductivity ($\gamma$) is calculated by the ratio between the photocurrent and the current at dark condition: $\gamma = (I_{photo}-I_{dark})/I_{dark}$. As it can be observed in Fig. 3(b), its maximum is reached at the gate voltage corresponding to the dark $V_{th}$. This shows that the device conductance is most photosensitive when the intrinsic free charges are depleted by the application of gate voltage, as it was previously mentioned[23]. Therefore, the relative photoconductivity acts as a compromise parameter that limits the $Q$ factor, since it decreases when the selectivity of the device is improved.

**IV. DISCUSSIONS**

It is well know that the surface is of paramount importance in controlling the electronic properties of ZnO nanowires. For instance, Soci *et al.*[16] described in detail that surface states are responsible for an energy band distortion (see Fig.3(c)) and the formation of an internal electric field. Consequently, electron-hole pairs are broken apart where electrons are forced to move to the central region of the NW, while holes are driven to the surface and get trapped[16,34]. As a result, there is an enormous increase of free carriers at the conduction band which are responsible for the photocurrent. These carries have long life times, due to the absence of available holes for recombination[16,20]. A similar analysis explains why the gate voltage can control the relative photoconductivity and quality factor of the photodetector. For a given $V_g$, an electric field is created in the nanowire, and this electric field bends the energy band up or down. Fig. 3(d) shows the energy band for a negative applied gate voltage (purple region). In this condition there is an electric field coming from the ZnO external surface to the Si substrate. This causes the bending of the energy band up, decreasing the number of thermally activated electrons. For a certain condition of negative gate voltage one can completely deplete the nanowire of carriers. Fig. 3(e) shows the opposite behavior for a positive applied gate voltage. In this case there is an electric field coming from the silicon substrate to the nanowire which bends the energy band down, increasing the carrier concentration and consequently the current in the nanowire. This process can happen whether the device is at dark conditions or under UV illumination and is the main reason why the gate voltage can be used to turn the photodetector off while still under illumination. Also, to make the photodetector more selective one can set the $V_g$ as close as possible to the threshold voltage due to the ZnO band gap, where the photo-absorption is largest. This will prevent creation of photocurrent for other wavelengths making the photodetector more selective. However, as the selectivity of the photodetector is improved there is a drawback related to the photodetector photosensitivity. For instance, the maximum photosensitivity is obtained for $V_{th}$ at dark conditions since this is the state where the maximum amount of carriers is generated from photoabsorption. Thus a decrease of the photodetector photosensitivity is expected for threshold voltages smaller or larger than the dark $V_{th}$.

## V. CONCLUSIONS

In summary we described the fabrication of a tunable ultra-violet detector using ohmic ZnO nanowire phototransistors. By the use of a gate voltage, full control of the photosensitivity and quality factor of the photodetection was achieved. This control is highly desirable for future applications where more precise and selective detection is required. One example would be the development of NW devices for the detection of airborne hazardous molecules or any molecule which has a discrete light emission.

## ACKNOWLEDGMENTS

We would like to thank Javier D. Sanchez-Yamagishi for helping with the manuscript and Prof. F. Plentz and Prof. P.S.S. Guimarães for helping with the optical setup. We acknowledge the financial support of CNPq/MCT, CAPES, and Fapemig. R. G. Lacerda is a CNPq fellow.

## REFERENCES

[1] P. Fei, P.H. Yeh, J. Zhou, S. Xu, Y.F. Gao, J.H. Song, Y.D. Gu, Y.Y. Huang, and Z.L. Wang, Nano Letters **9**, 3435-3439 (2009).

[2] H.J. Fan, W. Lee, R. Hauschild, M. Alexe, G. Le Rhun, R. Scholz, A. Dadgar, K. Nielsch, H. Kalt, A. Krost, M. Zacharias, and U. Gosele, Small **2**, 561-568 (2006).

[3] X.Y. Kong and Z.L. Wang, Nano Letters **3**, 1625-1631 (2003).

[4] X.D. Wang, J. Zhou, J.H. Song, J. Liu, N.S. Xu, and Z.L. Wang, Nano Letters **6**, 2768-2772 (2006).

[5] Q. Wan, Q.H. Li, Y.J. Chen, T.H. Wang, X.L. He, J.P. Li, and C.L. Lin, Applied Physics Letters **84**, 3654-3656 (2004).

[6] Q.H. Li, Y.X. Liang, Q. Wan, and T.H. Wang, Applied Physics Letters **85**, 6389-6391 (2004).

[7] Z.Y. Fan and J.G. Lu, Applied Physics Letters **86**, 123510 (2005).

[8] X.H. Wang, J. Zhang, and Z.Q. Zhu, Applied Surface Science **252**, 2404-2411 (2006).

[9] Z.Y. Fan, D.W. Wang, P.-C.C. Chang, W.-Y.Y. Tseng, and J.G. Lu, Applied Physics Letters **85**, 5923-5925 (2004).

[10] J. Yoon, W.K. Hong, M. Jo, G. Jo, M. Choe, W. Park, J.I. Sohn, S. Nedic, H. Hwang, M.E. Welland, and T. Lee, Acs Nano **5**, 558-564 (2011).

[11] Z.M. Liao, H.Z. Zhang, and D.P. Yu, Applied Physics Letters **97**, 033113 (2010).

[12] S.N. Das, K.J. Moon, J.P. Kar, J.H. Choi, J. Xiong, T.I. Lee, and J.M. Myoung, Applied Physics Letters **97**, 022103 (2010).

[13] Y.Q. Bie, Z.M. Liao, P.W. Wang, Y.B. Zhou, X.B. Han, Y. Ye, Q. Zhao, X.S. Wu, L. Dai, J. Xu, L.W. Sang, J.J. Deng, K. Laurent, Y. Leprince-Wang, and D.P. Yu, Advanced Materials **22**, 4284 (2010).

[14] A. Zhang, H. Kim, J. Cheng, and Y.H. Lo, Nano Letters **10**, 2117-2120 (2010).

[15] W.K. Hong, G. Jo, J.I. Sohn, W. Park, M. Choe, G. Wang, Y.H. Kahng, M.E. Welland, and T. Lee, Acs Nano **4**, 811-818 (2010).

[16] C. Soci, A. Zhang, B. Xiang, S.A. Dayeh, D.P.R. Aplin, J. Park, X.Y. Bao, Y.H. Lo, and D. Wang, Nano Letters **7**, 1003-1009 (2007).

[17] U. Ozgur, Y.I. Alivov, C. Liu, A. Teke, M.A. Reshchikov, S. Dogan, V. Avrutin, S.-J.J. Cho, H. Morkoc, U. Özgür, S. Doğan, and H. Morkoç, Journal of Applied Physics **98**, 41301 (2005).

[18] S. Xu, C. Xu, Y. Liu, Y. Hu, R. Yang, Q. Yang, J.-H. Ryou, H.J. Kim, Z. Lochner, S. Choi, R. Dupuis, and Z.L. Wang, Advanced Materials (Deerfield Beach, Fla.) **22**, 4749-53 (2010).

[19] C.S. Lao, M.-C.C. Park, Q. Kuang, Y.L. Deng, A.K. Sood, D.L. Polla, and Z.L. Wang, Journal of the American Chemical Society **129**, 12096-7 (2007).

[20] J. Zhou, Y.D. Gu, Y.F. Hu, W.J. Mai, P.-H.H. Yeh, G. Bao, A.K. Sood, D.L. Polla, and Z.L. Wang, Applied Physics Letters **94**, 191103 (2009).

[21] Z.-M.M. Liao, C. Hou, Q. Zhao, L.-P.P. Liu, and D.-P.P. Yu, Applied Physics Letters **95**, 093104 (2009).

[22] L. Polenta, M. Rossi, A. Cavallini, R. Calarco, M. Marso, R. Meijers, T. Richter, T. Stoica, H. Lüth, and H. Luth, ACS Nano **2**, 287-292 (2008).

[23] W. Kim and K.S. Chu, Physica Status Solidi (a) **206**, 179-182 (2009).

[24] Z.Y. Fan, P.-chun C. Chang, J.G. Lu, E.C. Walter, R.M. Penner, C.-hung H. Lin, and H.P. Lee, Applied Physics Letters **85**, 6128 (2004).

[25] Y.L. Pan, J. Hartings, R.G. Pinnick, S.C. Hill, J. Halverson, and R.K. Chang, Aerosol Science and Technology **37**, 628-639 (2003).

[26] H.C. Huang, Y.-L.L. Pan, S.C. Hill, R.G. Pinnick, and R.K. Chang, Optics Express **16**, 16523-16528 (2008).

[27] M. Tonezzer and R.G.G. Lacerda, Sensors and Actuators B-Chemical **150**, 517-522 (2010).

[28] S.M. Sze and K.K. NG, *Physics of Semiconductor Devices*, Third (John Wiley & Sons, Hoboken, 2006).

[29] J. Salfi, U. Philipose, S. Aouba, S.V. Nair, and H.E. Ruda, Applied Physics Letters **90**, 032104 (2007).

[30] S.-J.J. Chang, T.-J.J. Hsueh, C.-L.L. Hsu, Y.-R.R. Lin, I.-C.C. Chen, and B.-R.R. Huang, Nanotechnology **19**, 95505 (2008).

[31] J. Goldberger, D.J. Sirbuly, M. Law, and P. Yang, Journal of Physical Chemistry B **109**, 9-14 (2005).

[32] W.I. Park, J.S. Kim, G.-C.C. Yi, M.H. Bae, and H.-J.J. Lee, Applied Physics Letters **85**, 5052-5054 (2004).

[33] J. Maeng, G. Jo, S.-S.S. Kwon, S. Song, J. Seo, S.-J.J. Kang, D.-Y.Y. Kim, and T. Lee, Applied Physics Letters **92**, 233120 (2008).

[34] C.-J.J. Kim, H.-S.S. Lee, Y.-J.J. Cho, K. Kang, and M.-H.H. Jo, Nano Letters **10**, 2043-2048 (2010).

[35] M. Law, H. Kind, B. Messer, F. Kim, and P.D. Yang, Angewandte Chemie-International Edition **41**, 2405-2408 (2002).

[36] S. Song, W.-K.K. Hong, S.-S.S. Kwon, and T. Lee, Applied Physics Letters **92**, 263109 (2008).

[37] K. Huang, Q. Zhang, F. Yang, and D. He, Nano Research **3**, 281-287 (2010).